# Robust Monitoring of Arc Welding Processes: A Generalizable Framework with DVAE and Particle Filter

Yue Cao[1], Hai Lin[2*], YuMing Zhang[1*]

1: Department of Electrical and Computer Engineering and Institute for Sustainable Manufacturing, University of Kentucky, Lexington, KY 40506, USA

2: Department of Electrical Engineering, University of Notre Dame, IN 46556, USA

*: hlin1@nd.edu, yuming.zhang@uky.edu

**Abstract**: Arc welding processes are vital for continuous fabrication but are susceptible to disturbances that cause defects and compromise weld quality. Real-time monitoring is therefore essential, yet remains challenging due to complex visual patterns and the nonlinear, time-varying nature of welding dynamics. While deep learning offers potential, its reliance on large, labeled datasets, and in particular, on process- and application-specific tuning, limits industrial scalability. We question whether there exists a common approach to characterize major arc processes across different applications. If so, by observing and monitoring characteristic variables as process states under a unified framework, scalability can be greatly improved. This paper presents a robust monitoring framework generalizable across arc welding processes. It integrates deep latent representation learning to extract compact features from weld pool views in an unsupervised manner and employs Bayesian filtering to enhance robustness against sensory disturbances that are both sustained and fluctuating in arc welding, such as arc radiation and specular reflection from the weld pool surface. To this end, we employ a Dynamic Variational Autoencoder (DVAE), composed of a Convolutional Neural Network (CNN)-based encoder-decoder and a Long Short-Term Memory (LSTM)-based transition model, to jointly learn compact latent representations of weld pool images and their evolution under control inputs. This setup fuses instantaneous visual representations with process dynamics modeling, enabling compact latent features that satisfy both objectives. To achieve robust real-time inference, a specialized Particle Filter (PF) is introduced to jointly propagate the latent state and the hidden state of the LSTM transition model, preserving both current and historical process information while suppressing sensor disturbances such as arc rotation and specular reflection. This design is well suited to the relatively slow and inertial dynamics of arc welding, allowing the PF to effectively fuse model-based predictions with real-time observations. The proposed framework is validated on both GTAW and GMAW processes without process-specific customization, demonstrating its generalizability and robustness.

## 1. Introduction

### 1.1 Weld Penetration Sensing

Robotic welding is a key sector in industrial automation, offering high productivity, consistent quality, and reduced human labor and safety risks. Most welding robots, however, are still deployed for repetitive tasks such as spot welding with fixed parameters and predefined trajectories. By contrast, continuous fabrication in industries such as automobile manufacturing, shipbuilding, aerospace, and nuclear power relies heavily on arc welding processes, including Gas Tungsten Arc Welding (GTAW) and Gas Metal Arc Welding (GMAW). Nevertheless, adaptability to disturbances in continuous fabrication remains limited, leaving the automation of arc welding lagging behind.

In arc welding, an electric arc is established between a tungsten electrode or consumable wire and the workpiece, continuously melting and fusing the metallic materials to form a consistent weld seam [1]. Unpredictable process disturbances, such as welding parameter fluctuation, material inconsistency, and changes in thermal boundary conditions, significantly impact the resultant welding outcome [2]. Consequently, a critical challenge is to estimate the welding state that can reflect critical penetration characteristics from in-process sensory cues, providing the basis for real-time feedback control of arc welding.

The central task of arc welding is to achieve metallurgical bonding through adequate penetration, i.e., the extent to which molten metal fuses through the workpiece thickness. The penetration state can be described by quantifiable attributes such as penetration completeness, penetration depth, backside bead width, and the presence of lack-of-fusion defects [3]. However, direct online measurement is infeasible because penetration forms beneath the top surface, and the backside is mostly inaccessible due to fixture constraints.

Most existing estimation methods are, therefore, indirect and have focused on correlating measurable process phenomena, especially the weld pool, with penetration states, such as pool view [4], 3D surface profiles [5, 6], weld pool oscillation [7, 8, 9], acoustic signals [10], and infrared emissions [11]. The weld pool, where melting and solidification dynamics originate, contains most of the raw information related to weld penetration [12]. In [4], dynamic weld pool images captured by high dynamic range (HDR) cameras were used to predict backside bead width. To obtain the surface depth information that is more correlated with weld penetration, complex vision systems that leverage specular reflection of laser stripes on the pool surface for surface reconstruction [5], and a biprism-based stereo system that generates left and right image pairs for depth reconstruction [6], have been developed. Weld pool oscillation provides another sensing mechanism, as oscillation frequency decreases with increasing pool mass and inertia [7]. The reflected pattern of laser stripes reveals oscillation behavior [8], while arc voltage signals in pulsed GMAW can also indicate pool surface fluctuations [9]. Acoustic emission offers unique advantages by detecting subsurface penetration events, with time and frequency domain analysis effectively inferring penetration states [10]. Infrared sensing captures thermal radiation from weld pool surfaces, using emissivity differences between solid and liquid phases for boundary detection while mitigating arc light interference [11]. Collectively, these sensing modalities provide a rich, multi-dimensional data stream.

The vision-based monitoring, as the mainstream approach in molten pool sensing, has seen notable innovations in system design. For example, in Wire Arc Additive Manufacturing (WAAM), a camera system with narrow band optical filters, functioning as a quotient pyrometer, enables spatially resolved recording of the weld pool temperature distribution [13]. In addition, a ratio pyrometer integrates optical filters with bandwidths adapted to the arc spectrum, enabling nearly complete suppression of plasma radiation and allowing clear measurement of molten pool temperature and volume [14]. Two-color thermography has been shown effective for real-time melt-pool temperature and hardness prediction in GMAW, using only a commercial color camera [15]. A stereo vision system, with two HDR cameras positioned before and behind the welding torch, enables monitoring of the three dimensional geometric features of the weld pool [16]. To capture complex molten pool behavior in laser welding, such as keyhole oscillation, a vision based sensing system achieves very high frame rates, up to 4800 frames per second, using narrow band optical filters to suppress arc radiation and coaxial illumination to enhance molten pool visibility [17]. Further, for challenging monitoring scenarios such as climbing helium GTAW, where the weld pool position and viewing angles vary continuously and the morphology changes drastically, a high speed camera system is employed [18].

## 1.2 Deep Learning for Weld Penetration Inference

### 1.2.1 Supervised Learning

With the selected sensing method providing raw information, the next step is to map such sensory data to the weld penetration state. Earlier studies relied on handcrafted features informed by domain knowledge, such as geometric characteristics of the three-dimensional weld pool surface [5], the amplitude and oscillation frequency of reflected laser patterns [8], or characteristic arc voltage signals [9], and then fit the relationship between the selected features and the target penetration state based on experimental results. However, this approach is highly dependent on expert intuition in feature selection, which restricts its scalability across processes.

The advent of deep learning has shifted this paradigm. Instead of manually designing features, models are trained to directly learn the mapping from raw sensory data to penetration state, thereby improving generalization [19]. Supervised learning, with models trained on labeled datasets to predict penetration depth, state class, or bead width, has been most widely studied [3]. For example, convolutional neural networks (CNNs) have been developed to classify penetration into categories such as partial, moderate, full, and excessive, using weld pool images centered on the keyhole region [20]. Another study proposed a dual input CNN that combines weld pool images with photodiode signals, reporting improved accuracy compared to image-only approaches [21]. More advanced models, such as vision transformers, which capture global contextual relationships through attention mechanisms, achieve lower misclassification and better generalization in the prediction of GTAW penetration with four penetration categories [22]. Recognizing the dynamic and inertial nature of welding, researchers have extended static CNNs to spatiotemporal models, acknowledging that precise weld penetration prediction necessitates the use of historical information. To incorporate these temporal dependencies, time series neural networks such as Recurrent Neural Networks (RNNs) and their advanced variants, Long Short-Term Memory (LSTM) and Gated Recurrent Unit (GRU) networks, have been employed. For instance, CNN LSTM architectures that process sequences of weld pool images improved backside bead width prediction by leveraging temporal cues [4]. Other similar works that predict penetration from dynamic weld pool images include the use of LSTM with segmentation network structures [23] and dual input CNNs with self-attention [24]. Moreover, a Generative Adversarial Network (GAN) with GRU structure verifies the necessity of incorporating historical information for penetration prediction by synthesizing complex topside weld pool images from relatively simple backside images and welding parameter sequences [25].

Deep learning-based in-situ monitoring has also been extended to WAAM, which employs wire-and-arc processes for layer-by-layer deposition and requires strict control of geometry and defect formation. For example, CNN-based methods have been used to convert time-series voltage signals into image representations for real-time anomaly detection in WAAM builds, demonstrating effective identification of abnormal conditions [26]. Federated learning frameworks incorporating LSTM classifiers further enable secure, decentralized anomaly detection across multiple WAAM units without sharing raw sensor data [27]. Deep vision models such as ResNet and YOLO have been applied to classify molten-pool states and detect defects including hump and drop phenomena, while introducing indicators for early warning and process stabilization [28]. Multi-modal ensemble learning approaches that integrate electrical, thermal, and visual data have also been developed to enhance robustness and support proactive defect prevention [29].

Despite these advances, current supervised learning studies remain constrained by their dependence on large labeled datasets and tight coupling to specific sensing setups. Data annotation for welding process monitoring is not only labor intensive but also requires substantial domain expertise. Tasks such as accurately segmenting molten-pool morphologies or classifying defects demand precise boundary delineation and a solid understanding of welding physics. In addition, state-of-the-art studies usually focus on customized model structures tailored to specific sensing modalities or systems, which requires high expertise, limiting generalizability. These challenges highlight the need for frameworks with reduced annotation dependence and broader adaptability across sensing modalities.

### 1.2.2 Learning-Based Latent Representation

To overcome these constraints, researchers have explored unsupervised methods that learn compact latent representations of raw sensory data. A representative framework is the Variational Autoencoder (VAE), which encodes observations into a structured latent distribution through an encoder–decoder architecture [30]. This yields a compact, disentangled representation where latent components can capture distinct sources of variation, thereby enhancing interpretability. Variants like Vector-Quantized VAE (VQ-VAE) enhance structural consistency using discrete latent tokens [31], while the β-TC VAE promotes disentanglement through a total correlation penalty [32]. These advances make them particularly well-suited for complex welding sensory patterns. For penetration prediction, a VQ-VAE discretized weld pool images into structured latent tokens, which are then modeled with Transformer networks to capture temporal dynamics and predict penetration states such as insufficient, moderate, or excessive [33]. In our related work, a β-TC VAE was applied to obtain a disentangled latent representation of weld pool views, with temporal dependencies modeled through an LSTM-based regression network to enable dynamic image-based modeling of the GTAW process [34]. For anomaly detection in WAAM, a VQ-VAE encoded video streams into a structured latent space, where anomalies such as spatter, holes, and geometric defects were detected via reconstruction errors and outlier scores [35]. These studies demonstrate that latent representations can automatically serve as process states, offering interpretability and compactness.

However, a primary limitation of these approaches is the use of a static VAE, which encodes frames independently and disregard temporal dependencies inherent in welding. The learned representation is thus prone to transient sensory fluctuations rather than coherent states reflecting the inertial and dynamic nature of the process. Furthermore, even in [34] where temporal dynamics are modeled, the dynamic model is trained afterward on latent codes obtained with a static VAE. This decoupled design burdens the temporal model with learning both true process dynamics and compensating for inconsistencies embedded in the latent space by the static encoder. Such separation of spatial and temporal learning undermines stability and fidelity of the learned state. Therefore, a unified framework that jointly learns latent representations and process dynamics is needed to ensure consistent, temporally coherent state estimation.

## 1.3 Dynamic State Representation

To overcome the limitations of static VAEs, we propose a Dynamic Variational Autoencoder (DVAE) to learn latent states that capture the temporal dynamics of arc welding. Unlike standard VAEs that encode frames independently, a DVAE introduces a recurrent transition model (e.g., LSTM or GRU) that conditions each latent state on its predecessor, effectively structuring the latent space as a nonlinear state-space model [36]. This integration enforces both accurate reconstruction and coherent latent trajectories, encouraging separation of persistent process states from transient fluctuations.

DVAEs have been successfully applied across various domains for learning temporally consistent latent states. In chemical process monitoring, a DVAE represents time-sequential chemical process variables and quality measurements in a dynamic latent space, enabling reconstruction of missing quality measurements when only process variables are available [37]. For robotic movement learning, DVAE embeds high-dimensional human motion and robot joint sequences into compact latent spaces, enabling the generation of smooth robot movements [38]. For complex sequential data, hierarchical DVAEs that incorporate both static and dynamic latent variables to capture time-invariant and time-dependent information, have been applied human motion analysis [39] and speech processing [40].

However, despite these advances, the application of DVAEs to complex imaging-based manufacturing processes such as arc welding remains notably limited. A critical challenge lies in the data complexity: prior applications often relied on relatively low-dimensional streams with clear physical meaning, such as process variables, robot joint angles, or speech spectra, whereas welding image sequences are high-dimensional and semantically complex. As a result, a DVAE for arc welding must not only capture temporal dependencies but also extract and interpret rich semantic information from image data. This makes learning a robust temporal latent representation for welding significantly more challenging. In addition, the unpredictable disturbances in the welding process and imaging conditions introduce additional noise, making real-time inference even more difficult. The DVAE models the process state in its latent space and learns the process dynamics, but for real-time application, its predictions are susceptible to the unpredictable disturbances and imaging noise in welding, necessitating a robust inference mechanism.

To achieve robust online state estimation under such noise and complexity, we employ Bayesian filtering, a principled framework that is naturally suited to the state-space structure learned by the DVAE. Bayesian filters maintain a probabilistic belief over the system state and iteratively update it by fusing prior predictions from a process model with new observations [41]. Classic approaches, such as the Kalman Filter and its nonlinear extensions [41], have been applied to weld penetration estimation [9] and soft robot state estimation [42]. More recently, Deep Bayesian Filters [43] have extended this framework to situations where the dynamic model is learned through deep networks, such as DVAEs. They have been applied in domains with low-dimensional, physically clear states—for instance, in soft robotics, where actuator kinematics and sensor dynamics are integrated within a Bayesian framework for robust state estimation [44]; and in radar-based object tracking, where noisy position measurements are fused with estimated object shape extensions using an LSTM-enhanced filter to improve temporal consistency [45]. In contrast, arc welding involves high-dimensional, semantically complex weld pool imagery, presenting a much greater challenge for Bayesian filtering.

## 1.4 Proposed Work

This work introduces an unsupervised probabilistic framework that integrates deep latent representation learning with Bayesian filtering for robust state estimation in arc welding. The main contributions are:

(1) A DVAE with a CNN based encoder decoder structure for semantic interpretation and compression, and an LSTM transition model to capture dynamics between latent evolution and welding parameters, is proposed to learn a temporally consistent latent space.

(2) A Particle Filter is specially designed for the LSTM represented inertial and slow changing welding latent dynamics, with LSTM cell internal states included in the particle set for complete state representation. We demonstrate that the inferred state estimation accurately reflects backside bead formation as a measure of weld penetration in GTAW.

(3) The framework demonstrates generalizability across arc welding processes, including GTAW and GMAW, without process specific customization, providing a scalable method for arc welding state inference.

Thus, our work proposes a generalizable state construction method for arc welding that learns temporally coherent latent representations directly from raw sensory data and enhances robustness through Bayesian inference. This approach establishes a new foundation for a unified, unsupervised, and physics aware paradigm in welding process monitoring.

The paper is organized as follows. Section 1 introduces the principles of arc welding and Bayesian Filter based welding state estimation. Section 2 presents the DVAE for dynamic welding state representation using a GTAW process. Section 3 details the Particle Filter algorithm and results. Section 4 verifies the probabilistic state estimation framework's generalizability using a GMAW process. Section 5 concludes the current work and discusses future directions.

## 2. Probabilistic Framework for Arc Welding State Estimation

In arc welding, the weld pool geometry dictates both the grain growth of the weld metal and the resulting penetration. Grain formation follows a nucleation and growth process: nucleation occurs at the solid–liquid interface, and grains then grow along the maximum temperature gradient, i.e., normal to the weld pool boundary [43]. As a result, weld penetration, as a reflection of the solidified molten metal, can be inferred from the topside weld pool geometry, verified by previous studies [4-6, 20-24]. In this study, we therefore construct the process state capable of reflecting crucial process information including penetration from dynamic weld pool views.

In addition, the weld pool geometry, which is primarily dictated by key welding parameters, exhibits limited variations. Process parameters such as current and welding speed determine the forces and heat input in the process, which in turn shape the weld pool appearance through coupled heat transfer and fluid flow mechanisms [46]. Since the ways in which input parameters can change are limited in dimension, the weld pool geometry, typically expressed in terms of area, curvature, and brightness distribution as observed in 2D images, also shows limited patterns. This has been further verified through dimensionless analysis, which demonstrates that weld pool geometrical variables can be expressed by dimensionless combinations of process parameters and material properties. For example, it has been shown that wavy weld pool fusion boundaries can be derived from such dimensionless groups [47], while two dimensionless numbers for weld pool width and aspect ratio have been identified from experimental data using a data-driven framework [48]. However, these strict relationships are obtained from numerical simulations or fixed experimental conditions, serving as explanatory tools rather than field monitoring.

Building on these foundations, it is reasonable to represent weld pool images in a structured low-dimensional latent space. This can be achieved in an automatic, data-driven manner using variational inference models, as will be detailed later. Accordingly, this work defines the latent process state $x_t$ as the compact representation of the weld pool at time $t$. The state $x_t$ captures essential information about the weld pool morphology that is tightly correlated with penetration, while remaining low-dimensional to reflect the limited degrees of variation in pool geometry.

Due to the time-dependent and spatially distributed nature of heat transfer, the evolution of the weld pool depends not only on current inputs but also on past conditions, making welding inherently history-dependent. We adopt the following probabilistic formulation of the welding process:

$$x_t = p_{\theta_{trans}}(x_t | x_{t-1:t-n}, u_{t-1:t-n}) \tag{1}$$

$$z_t = p_{\theta_{obs}}(z_t | x_t) \tag{2}$$

where $u_t$ is the process input (e.g., welding current, speed), $x_t$ is the latent process state, and $z_t$ is the observation (e.g., weld pool image). The transition model $p_{\theta_{trans}}$ captures the process dynamics under control inputs, while the observation model $p_{\theta_{obs}}$ captures the mapping from latent states to observed images, parameterized by $\theta_{trans}$ and $\theta_{obs}$, respectively. In this work, we learn $p_{\theta_{trans}}$ and $p_{\theta_{obs}}$ jointly within a Dynamic Variational Autoencoder, with a probabilistic LSTM network as the transition model, to ensure latent representation qualify for frame-wise representation and temporal consistency.

Due to the disturbances in the welding process and the presence of imaging noise, the obtained transition model and observation model are embedded within a Bayesian filtering framework for online state estimation:

$$p(x_t | z_{t-1:t-n}, u_{t-1:t-n}) \propto p_{\theta_{obs}}(z_t | x_t) \int p_{\theta_{trans}}(x_t | x_{t-1:t-n}, u_{t-1:t-n})\, p(x_{t-1} | z_{t-1:t-n}, u_{t-1:t-n}) dx_{t-1} \tag{3}$$

Here, the prior $p_{\theta_{trans}}(x_t | x_{t-1:t-n}, u_{t-1:t-n})$ provides model-based predictions of the next latent state, while the likelihood $p_{\theta_{obs}}(z_t | x_t)$ incorporates real-time sensory evidence from weld pool images. The recursive update produces the posterior $p(x_t | z_{t-1:t-n}, u_{t-1:t-n})$, which balances predictive dynamics with corrective measurements, yielding robust state estimates even under process disturbances and noisy observations. To implement this inference in practice, we employ a Particle Filter, which maintains a set of candidate latent states and is particularly well suited to handling the nonlinear dynamics represented by the LSTM transition model

# 3. Dynamic Variational Autoencoder

## 3.1 DVAE Architecture and Training

At the core of the proposed state estimation framework is the Dynamic Variational Autoencoder (DVAE), which learns a structured latent space for weld pool dynamics. As illustrated in Fig. 1, the DVAE consists of two modules: (1) a convolutional encoder–decoder for image representation learning, and (2) a recurrent transition model that captures temporal dependencies between latent states and welding parameters.

The encoder maps a weld pool image into a compact Gaussian distribution parameterized by mean $\mu_t$ and variance $\Sigma_t$. It is built on a deep convolutional backbone that progressively reduces spatial resolution while extracting hierarchical features. A Convolutional Block Attention Module (CBAM) is integrated to apply both channel and spatial attention, emphasizing regions of the weld pool most relevant to penetration and bead formation. Along the encoding path, the tensor dimensions are (3,128,128), (32,64,64), (64,32,32), (128,16,16), and (256,8,8). The final feature map is flattened and passed through fully connected layers to produce the mean and variance vectors in a low-dimensional latent space (dimension 5 in this work). A latent welding state $x_t$ is then obtained by sampling from the Gaussian distribution $\mathcal{N}(\mu_t, \Sigma_t)$. The decoder mirrors the encoder with upsampling and deconvolution layers to reconstruct weld pool images from the latent state. Attention modules are also incorporated in the decoding path to refine reconstruction quality, ensuring both spatial dependencies and fine-grained details are effectively restored.

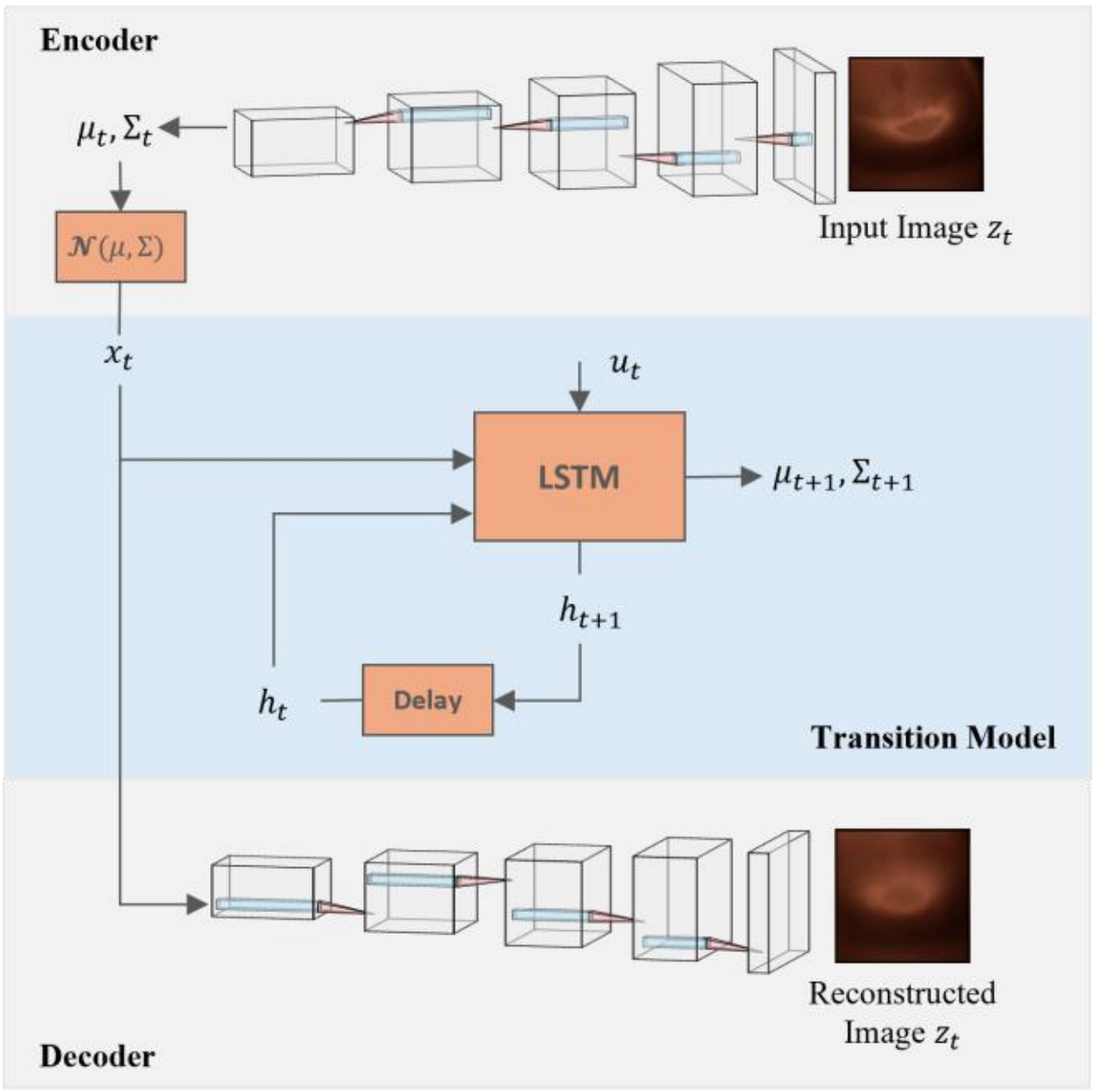

Fig.1. Structure of Dynamic Variational Autoencoder

The transition model, which represents the latent dynamics under welding parameters, is implemented using an LSTM cell. At each step, the LSTM receives as input the current latent state $x_t$, the process input $u_t$ (e.g., welding current or speed), and its internal memory consisting of the cell state $c_t$ and the hidden state $h_t$. Both internal states are 128-dimensional vectors. The cell state carries long-term information across time, while the hidden state encodes short-term contextual features; together they allow the LSTM to model both persistent and transient effects in sequential data [49]. The internal states are recursively updated and retained for the next step. Thus, the transition model can be expressed as:

$$\mu_{t+1}, \Sigma_{t+1} = p_{\theta_{trans}}(u_t, x_t, h_{t+1}, c_{t+1}) \quad (4)$$

This model's design reflects a structured approach to uncertainty in welding. While core physical dynamics (e.g., heat transfer, fluid dynamics) are predominantly deterministic, stochasticity arises from sensor noise and unmodeled effects. Our architecture partitions this uncertainty: the LSTM transition function learns a deterministic mapping representing the core dynamics, with its states $(h_t, c_t)$ serving as stable historical memory. Crucially, the LSTM operates on a sampled latent state $x_t \sim \mathcal{N}(\mu_t, \Sigma_t)$, rather than using $(\mu_t, \Sigma_t)$ as input. This sampling step enforces proper uncertainty propagation and avoids collapse into an overconfident deterministic model. In this way, the design provides robust predictions consistent with the generative semantics of the DVAE.

The DVAE is trained by optimizing the Evidence Lower Bound (ELBO), which jointly balances reconstruction fidelity and the consistency of latent dynamics. The ELBO is formulated as:

$$L(\theta,\phi) = E_{q_\phi(x_t|z_t)}\left[log\, p_{\theta_{dec}}(z_t|x_t)\right] - D_{KL}\left(q_\phi(x_t|z_t)||p_{\theta_{trans}}(x_t|u_{t-1},x_{t-1},h_{t-1},c_{t-1})\right) \tag{5}$$

Here, $q_\phi(x_t|z_t)$ is the approximate posterior from the encoder, $p_{\theta_{dec}}(z_t|x_t)$ is the decoder as the likelihood model, and $p_{\theta_{trans}}(x_t|u_{t-1},x_{t-1},h_{t-1},c_{t-1})$ is the transition prior defined by the LSTM dynamics. The KL divergence term ensures the encoder's posterior stays aligned with the predictions of the learned dynamics model. To ensure robust long-term prediction, which is essential for welding with slow and inertial dynamics, a specialized training strategy is employed, as illustrated in Fig. 2. For each experimental sequence of length $L$, the first half is used to initialize the LSTM's internal states $(h_t, c_t)$ using encoded latent states from the observation, thereby capturing historical context. The second half is then predicted autoregressively by the transition model using only the process inputs $u_t$, without observational correction. This forces the transition model to learn accurate long-horizon dynamics. Meanwhile, the encoder and decoder are trained to minimize reconstruction loss across the entire sequence, ensuring faithful per-frame representation. This approach yields a latent space that captures both immediate visual features and temporally consistent dynamic behavior.

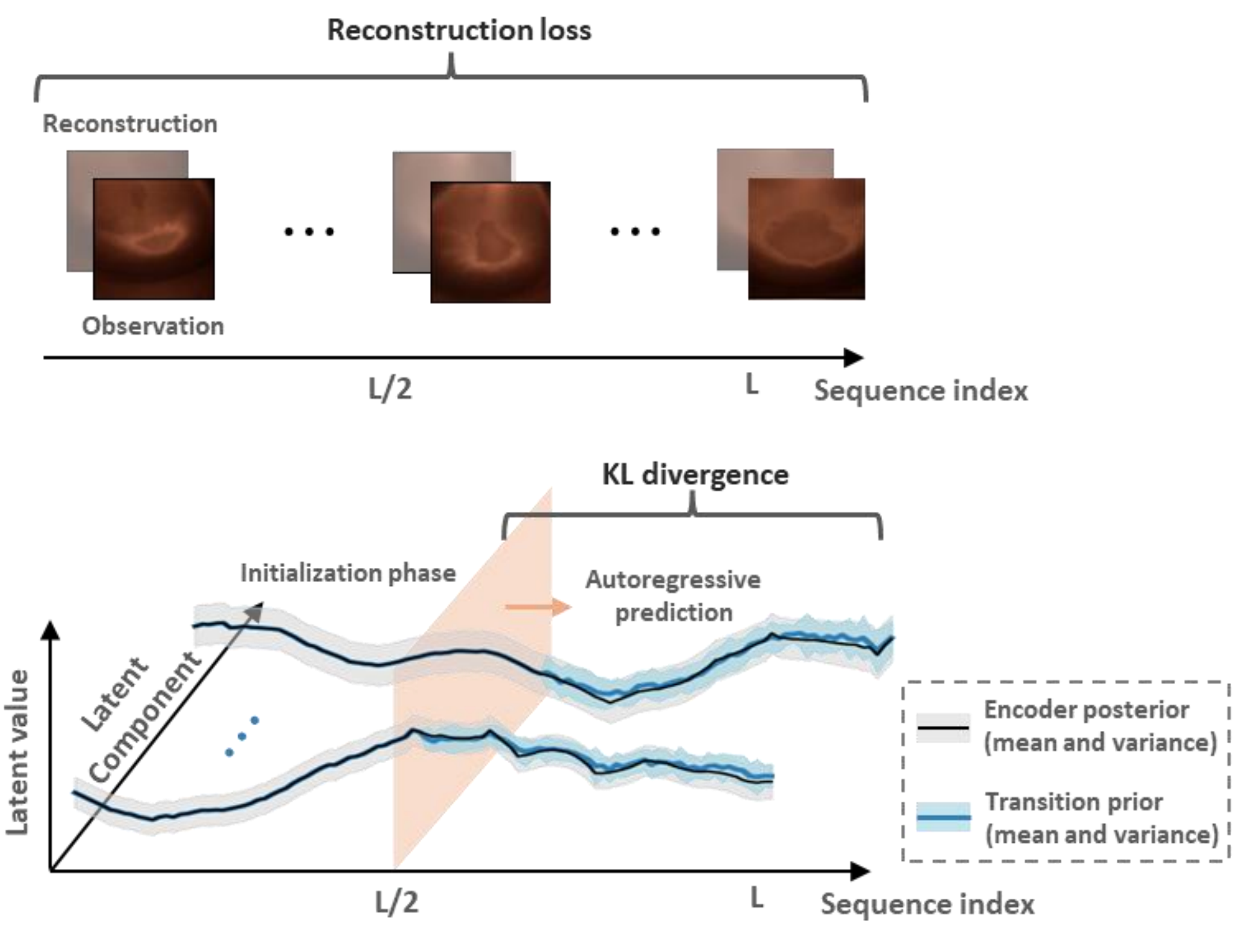

Fig. 2. ELBO Loss

## 3.2 GTAW Dataset

The proposed probabilistic state estimation framework is first validated in Gas Tungsten Arc Welding (GTAW), a widely used arc welding technique valued for its stable arc, precise heat input control, and clean welds. The experimental system is illustrated in Fig. 3, with welding parameters summarized in Table 1. The core setup is a robotic welding platform composed of a UR15e robot for torch manipulation, a Miller Dynasty 350 power source, and an XVC-1100 camera mounted on the robot arm to capture weld pool images from a relatively stationary viewpoint. The camera is positioned at a 45° angle to the workpiece and equipped with a band-pass filter to suppress arc-light interference. To ensure proper shielding, the weld zone is protected by top-side argon shielding from the torch and backside inert-gas purging through an enclosed back chamber. Welding is carried out under a pulsed-current mode: the peak current ensures sufficient penetration, while the base current sustains the arc. A pulse frequency of 10 Hz with a 50% duty cycle is applied, and weld pool images are captured during the base-current phase (one frame per cycle) to provide clear pool boundaries. In this system, the welding speed serves as the sole control input, which is adjusted in real time by a human operator via a virtual-reality-based teleoperation system [50].

Table 1. Welding Parameters for the GTAW Experiments.

| Welding Parameters | Values |
|---|---|
| Welding method | GTAW |
| Welding power | Miller Dynasty 350 |
| Polarity | DCEN |
| Welding speed | 2.5-5.5 mm/s |
| Material | Stainless steel 304 L |
| Thickness | 1.85 mm |
| Gas | 99.99% Ar |
| Shielding gas flow rates (top and backside) | 15 $ft^3/h$ |
| Arc length | 5 mm |
| Current frequency | 10 Hz |
| Peak current | 100 A |
| Base current | 20 A |
| Pulse duty cycle | 50% |

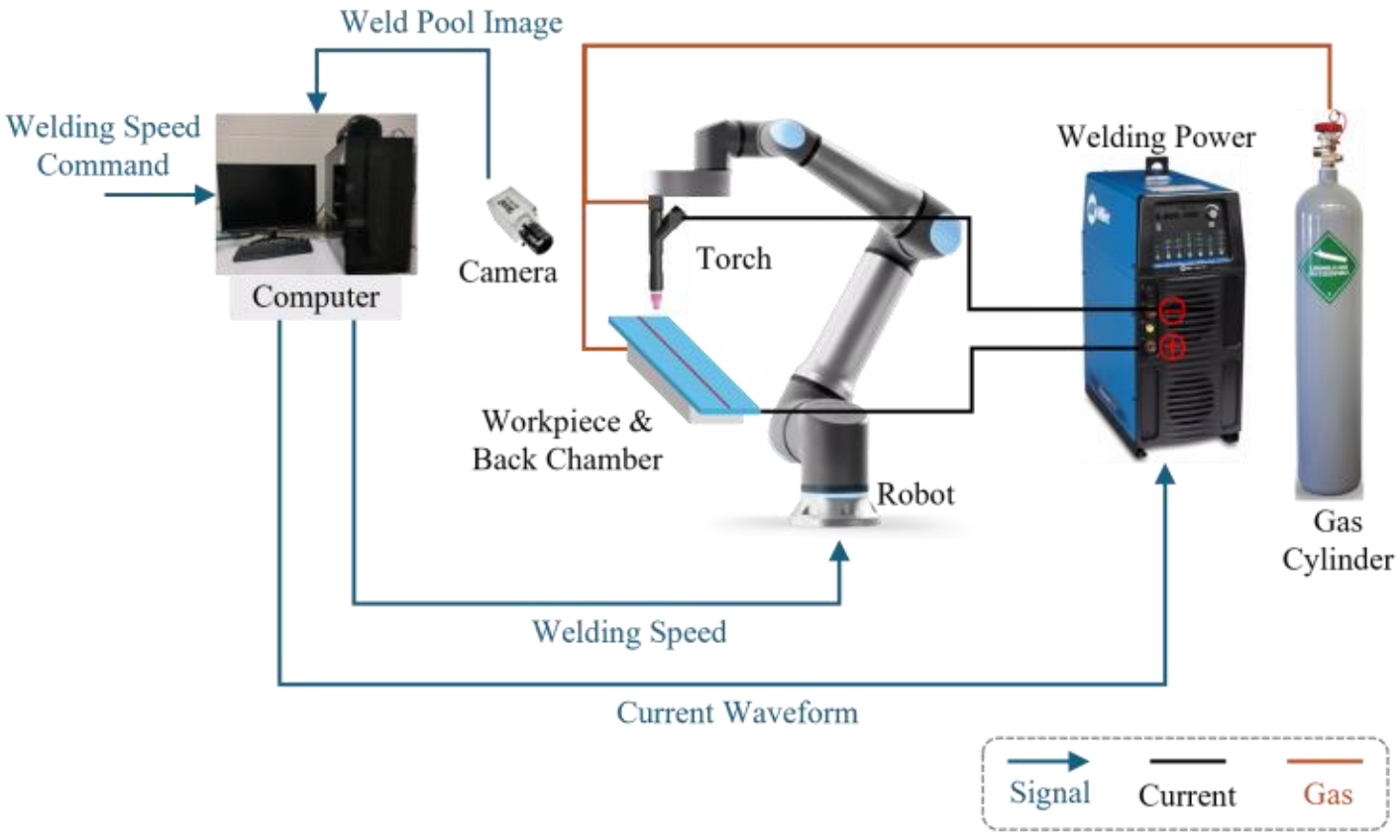

Fig. 3. GTAW Experimental System.

In the GTAW experiments, human-in-the-loop operation was employed to counteract deliberately introduced disturbances, thereby creating varying welding speed profiles to stimulate process dynamics. The backside bead width was measured after welding to evaluate whether the estimated states reflect the ground-truth penetration (Fig. 4, one experiment). Diverse weld pool variations under different heat input conditions were also observed (Fig. 5, one experiment). As observed, weld pool images vary in size, boundary contour, and brightness distribution depending on welding speed, external disturbances, arc light interference, and imaging conditions. A general trend is that weld pool size increases as welding speed decreases, since greater heat input melts more material and enlarges the pool. The DVAE is therefore designed to model these dynamic semantic changes linked to welding speed. A total of eight GTAW experiments were conducted, each lasting 70–110 s with a sampling rate of 10 Hz. From these raw sequences, training segments of length $L = 100$ image–speed pairs were extracted. The first 50 frames were used to initialize the LSTM internal states, and the remaining 50 frames were used for autoregressive prediction. Using a sliding window with a step size of 5, we obtained 555 training sequences for DVAE learning.

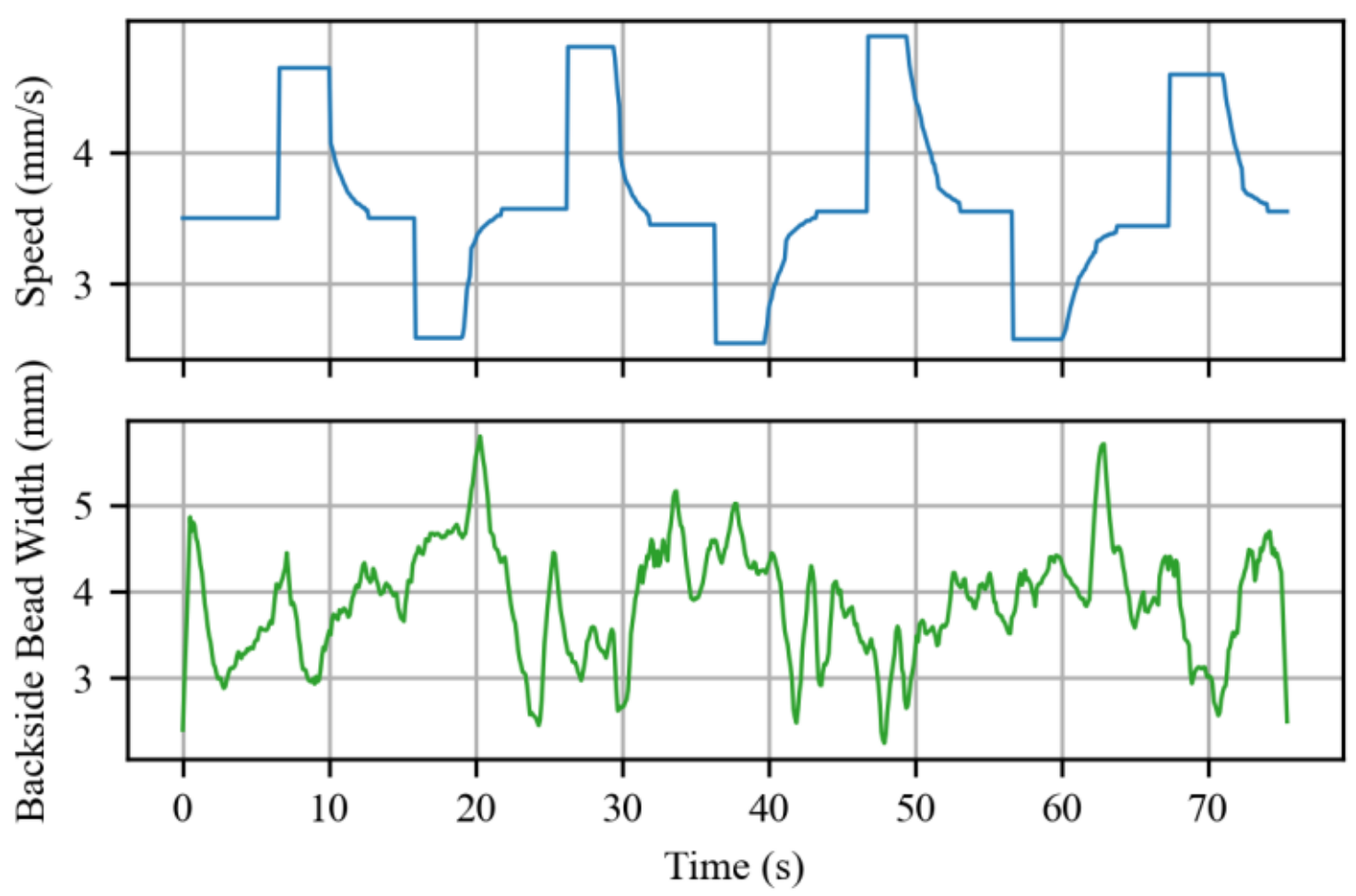

Fig. 4. Welding Speed and Backside Bead Width.

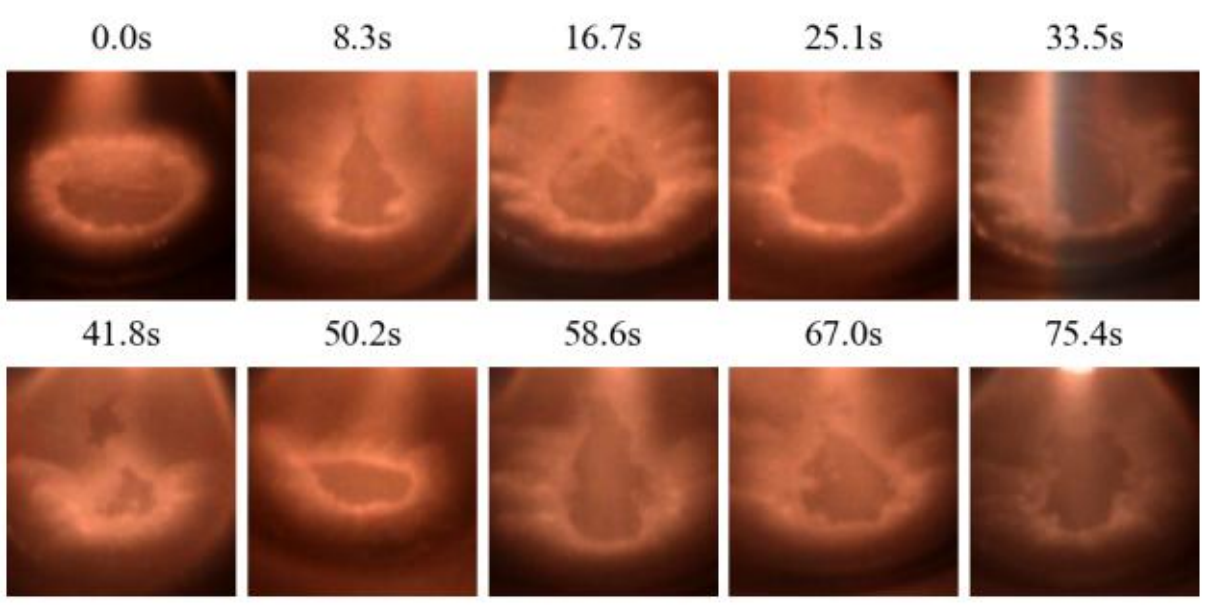

Fig. 5. Weld Pool Views.

### 3.3 DVAE Training Results

The DVAE was trained for 100 epochs on 555 sequences, each consisting of 100 frames of weld pool images paired with welding speed inputs. In each iteration, four sequences were sampled as a batch (400 frames in total). A two-optimizer scheme was adopted: the CNN encoder–decoder was updated with Adam (learning rate $1\times10^{-4}$) using 25 micro-batches of 16 images each, providing sufficient updates given its larger parameter count, while the LSTM transition model was updated once per outer batch with Adam (learning rate $1\times10^{-3}$). In both cases, updates were guided by the overall ELBO objective in Eq. 5, which balances reconstruction fidelity with a KL divergence term that aligns the encoder posterior with the transition prior. The training curves (Fig. 6) report the ELBO evaluated on each full batch, reflecting the joint optimization of the visual and temporal components.

Fig. 6 shows the evolution of the total DVAE loss together with the ratio between reconstruction and KL divergence terms. After an initial sharp adjustment, where the ratio drops rapidly, both terms gradually stabilize. The reconstruction term remains dominant, and the ratio converges toward near 20, indicating that the encoder–decoder captures most of the visual information, while the KL term provides steady regularization to align the learned dynamics. The smooth convergence of both curves demonstrates stable joint training of the visual and temporal components.

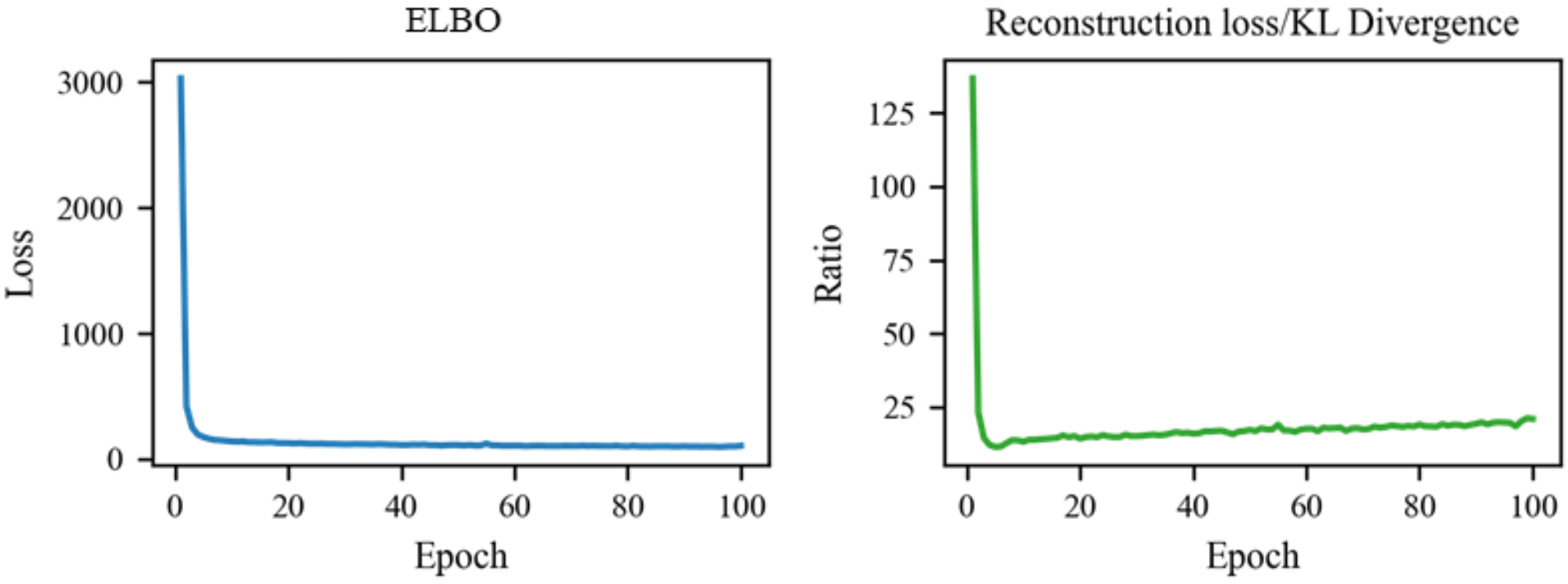

Fig. 6. Training Curves of DVAE on GTAW Dataset.

Fig. 7 presents the comparison between reconstructed weld pool images and the corresponding original images using the trained encoder-decoder. The reconstructions resemble the originals in overall size and boundary but appear more blurred, with subtle irregular details in the pool morphology are not preserved. This indicates that the learned latent representation successfully captures the essential geometry of the weld pool while filtering out stochastic, less significant variations.

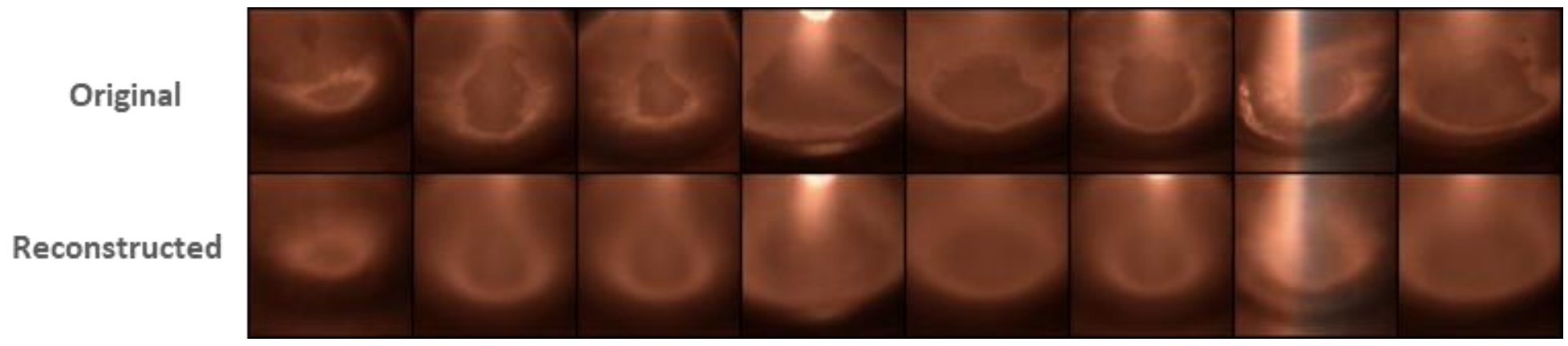

Fig. 7. Reconstruction of GTAW Weld Pool Images**.**

Fig. 8 illustrates the validation of the final transition model on a 500-step experiment sequence (0.1 s per step). The first 50 steps are used for initialization, during which the LSTM internal states $(h_t, c_t)$ are set with historical context. Thereafter, driven by the varying speed signal, the weld pool images evolve dynamically, leading to corresponding changes in the encoder posterior $q_\phi(x_t|z_t)$, represented as oscillating latent components with mean and variance bands (blue). In parallel, the transition prior $p_{\theta_{trans}}(x_t|u_{t-1}, x_{t-1}, h_{t-1}, c_{t-1})$, driven solely by the control input $u_t$ after initialization, produces predictions (pink) based on the learned dynamics. The transition prior responds consistently to changes in the speed signal, and its overall agreement with the encoder posterior, demonstrating the effectiveness of the learned LSTM transition model. Nonetheless, purely model-driven predictions cannot compensate for unpredictable disturbances, leading to error accumulation over time. This highlights the necessity of integrating both the transition prior and the encoder posterior for robust state estimation, as will be shown later.

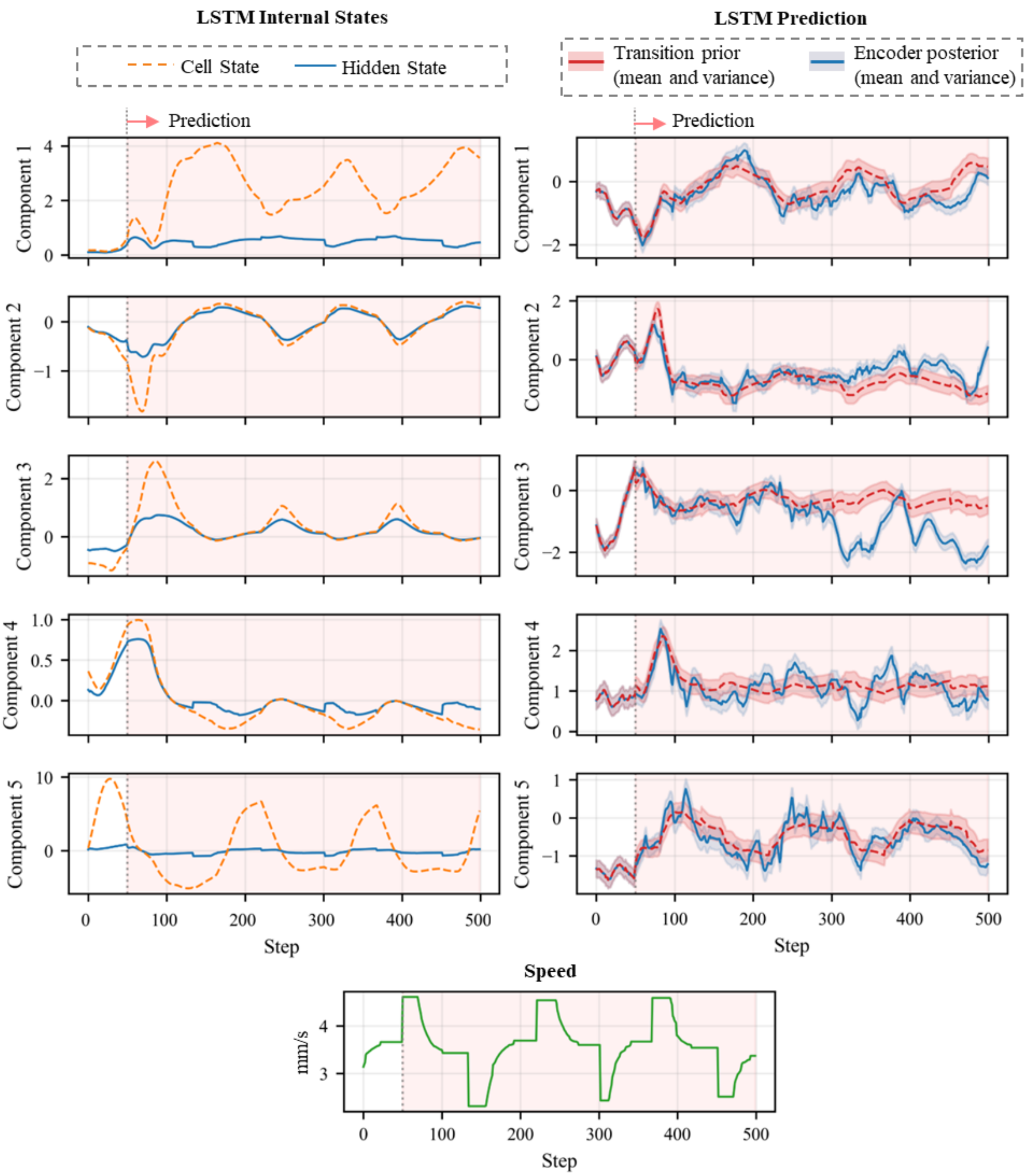

Fig. 8. Validation of LSTM Transition Model on GTAW Experiment.

## 4. Particle Filter for State Estimation

This section presents the design of a Particle Filter (PF) for Bayesian state estimation within the DVAE framework. The DVAE supplies two complementary sources of information: the encoder posterior $q_{\phi}(x_t|z_t)$, grounded in visual observations, and the transition prior $p_{\theta_{\text{trans}}}(x_t|u_{t-1}, x_{t-1}, h_{t-1}, c_{t-1})$, derived from learned dynamics. Individually, each is flawed: the prior drifts over long horizons due to accumulated error and model bias, while the posterior is sensitive to sensor noise. The PF fuses these signals in a recursive Bayesian manner, implementing a sequential Monte Carlo approximation of the posterior $p(x_t|z_{t-n:t-1}, u_{t-n:t-1})$. By propagating a set of weighted particles, it provides a robust probabilistic estimate of the latent welding state $x_t$ that remains both temporally consistent and observationally accurate, suppressing noise and mitigating drift.

The key innovation of our PF design lies in its ability to incorporate the DVAE's LSTM transition model without sampling the high-dimensional LSTM internal states. This enables the filter to preserve temporal consistency while remaining computationally tractable for real-time process monitoring. The technical implementation of the PF is provided in the Appendix.

The PF successfully integrates the dual estimates, as illustrated in Fig. 9. The encoder-only estimate is responsive but exhibits high-variance fluctuations due to observation noise and disturbances. Conversely, the transition-prior trajectory is smooth but prone to drift and can diverge from the true state over time. The PF output effectively reconciles these two sources: the particles are pulled toward the encoder's observation when it is reliable, providing correction, but their evolution is governed by the smooth dynamics of the transition model, ensuring temporal coherence. The resulting trajectory of $\hat{x}_t$ retains the smoothness of the model while being regularly corrected by the observations. The accompanying variance estimate $\mathrm{Var}[\hat{x}_t]$ reliably increases during periods of high dynamic uncertainty or poor observation quality, providing a principled confidence measure. This robust probabilistic fusion makes the PF estimate far more suitable for reliable process monitoring and closed-loop control than either source alone.

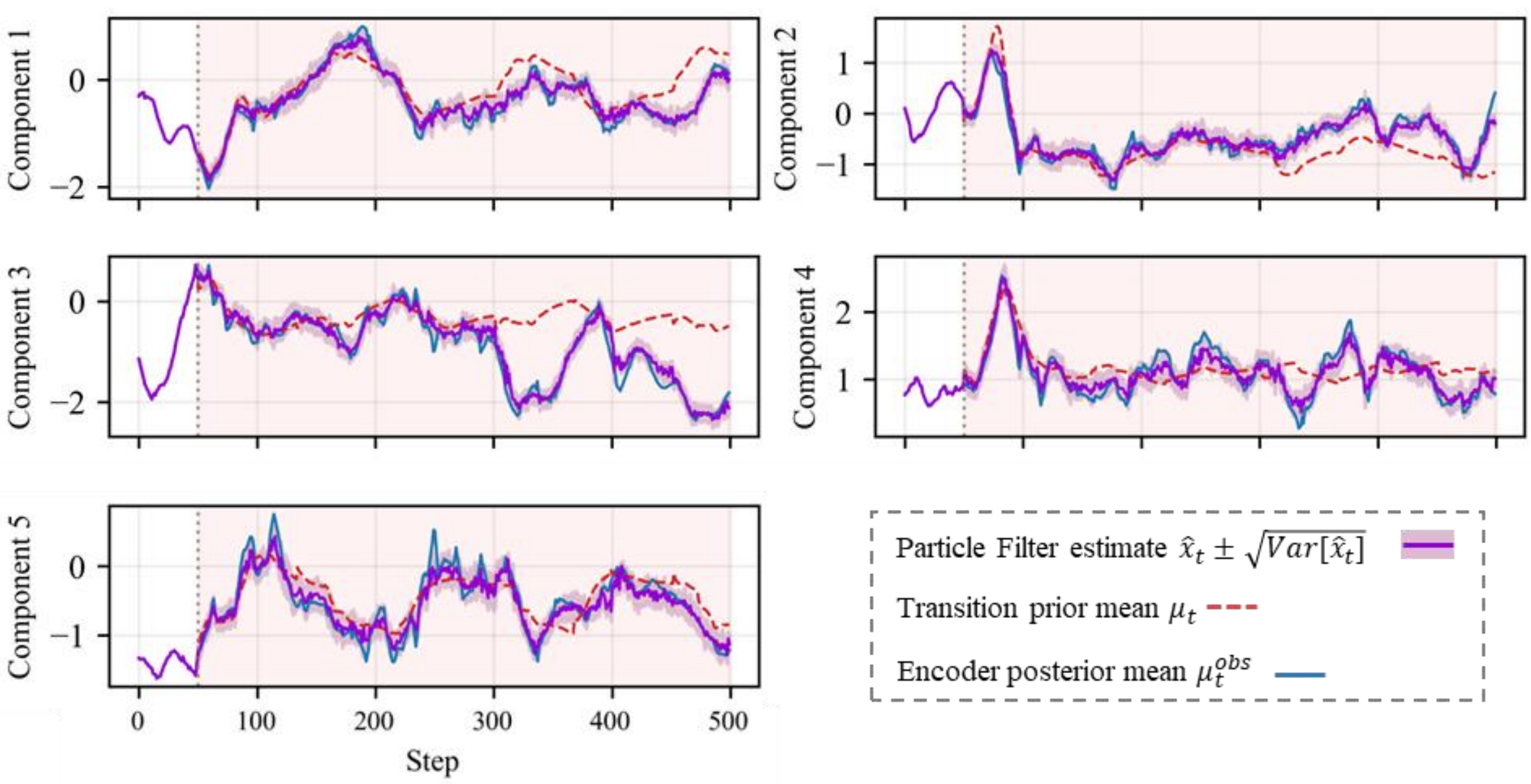

Fig. 9. Particle Filter State Estimation Results in GTAW.

The PF estimate $\hat{x}_t$ provides a smooth, robust latent representation of the welding process that can be exploited for downstream tasks such as weld penetration monitoring and process control. To illustrate this, weld penetration was quantified by correlating $\hat{x}_t \in \mathbb{R}^5$ with the backside bead width $W_b$. A quadratic regression model was identified using three of the eight GTAW experiments for training and validated on the remaining ones:

$$W_b = \beta_0 + \sum_{i=1}^{5} \beta_i \hat{x}_{t,i} + \sum_{i=1}^{5} \sum_{j=i}^{5} \beta_{ij}\, \hat{x}_{t,i} \hat{x}_{t,j}. \tag{6}$$

Figure 10 shows a representative case where the PF-based prediction overlaps closely with the measured bead width (RMSE $= 0.27$ mm). This indicates that $\hat{x}_t$ encodes physically meaningful process information; recovering penetration from $\hat{x}_t$ serves as concrete evidence that the learned state captures salient welding dynamics. As $\hat{x}_t$ is learned unsupervisedly from temporal sequences, so only a small labeled set is needed to calibrate $\hat{x}_t$ to $W_b$, yielding a data-efficient and sensor-robust monitoring pipeline. This contrasts with prior work [2], where deep networks were trained in a fully supervised manner to map raw images to penetration and thus required large labeled datasets.

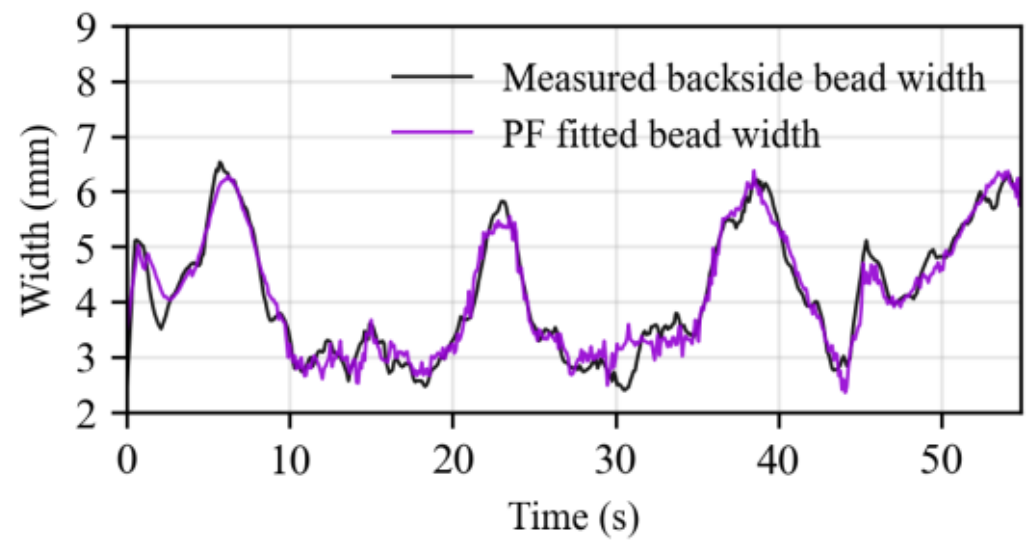

Fig. 10. Backside Bead Width Estimated from Particle Filter Results.

## 5. Generalization to Gas Metal Arc Welding

To evaluate the generalization capability of the proposed framework, we applied it to a bead-on-plate Gas Metal Arc Welding (GMAW) process. The experimental setup was identical as in Section 4.2. In this case, welding speed was the only parameter varied in real time, pre-programmed as a stochastic step signal (Fig. 11), while all other process parameters were held fixed, as shown in Table 2. As the heat input changed, the weld pool appearance varied accordingly (Fig. 12). To suppress the strong arc radiation in GMAW, a band-pass optical filter centered at 650 nm was employed, yielding high-contrast weld pool images.

Table 2. Welding parameters for the GMAW experiment.

| Welding Parameters | Values |
|---|---|
| Welding method | GMAW |
| Welding type | Bead on plate |
| Polarity | DCEP |
| Welding speed | 3–12 mm/s |
| Voltage | 40.8 V |
| Wire feed speed | 310 in./min |
| Shielding gas (95% Ar / 5% $CO_2$) | 15 $ft^3$/h |
| Contact tip–work distance | 20 mm |
| Wire diameter | 1.2 mm |
| Workpiece thickness | 20 mm |

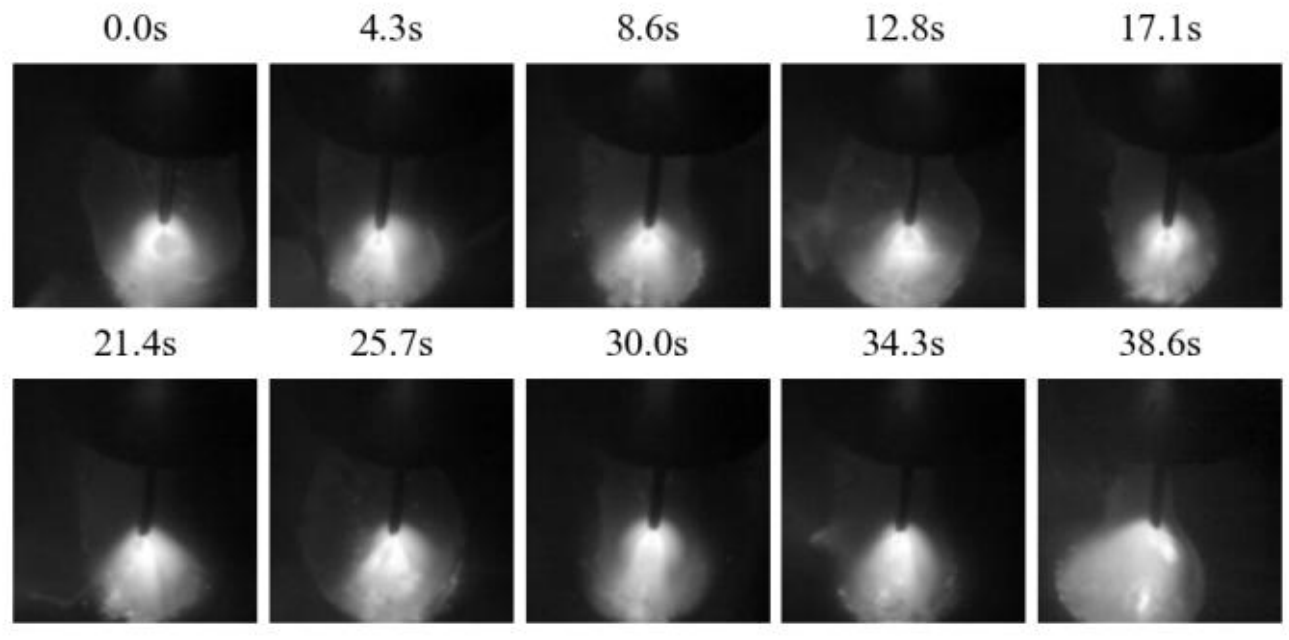

Fig. 12. GMAW weld pool views.

Compared to GTAW, obtaining a stable latent representation in GMAW is more challenging due to the increased process complexity. GMAW introduces additional process dynamics due to periodic molten droplet transfer from the wire to the pool. In addition, more intense arc radiation, fume, and larger temporal fluctuations in pool morphology bring stronger imaging disturbances. To verify generalizability, we evaluated the framework on GMAW using the same sensing setup and a comparable dataset volume (eight experiments, as in the GTAW study). The DVAE architecture, training hyperparameters, and PF settings were kept identical to those used for GTAW. Figure 13 shows DVAE reconstruction samples on a GMAW experiment. The decoder reconstruction preserve the principal pool characteristics, such as overall brightness distribution, weld pool boundary, and weld pool trailing region, showing that the learned latent representation retains the dominant semantics from image frames. Figure 14 presents the PF state estimate $\hat{x}_t$ alongside the encoder posterior mean $\mu_t^{\text{obs}}$ and the transition-prior mean $\mu_t$, with the speed input $u_t$ shown for reference. As expected, $\mu_t^{\text{obs}}$ exhibits high-frequency variability due to process noise, whereas $\mu_t^{\text{prior}}$ is smoother but can drift during long-horizon prediction. The PF fuses these complementary signals, yielding a state estimate that is both accurate and temporally coherent. Overall, these results demonstrate that the DVAE–PF pipeline generalizes effectively to GMAW. More broadly, the same framework can be applied to other arc welding processes with dynamic sensor data (image or alternative modalities) by learning a temporally consistent latent representation via the DVAE and improving inference robustness through particle filtering.

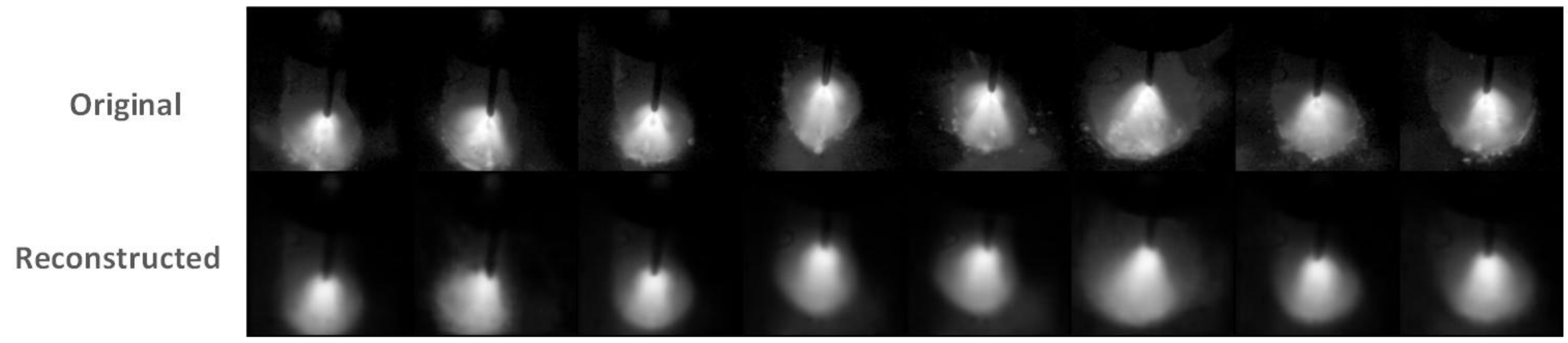

Fig. 13. Reconstruction of GMAW Weld Pool Images.

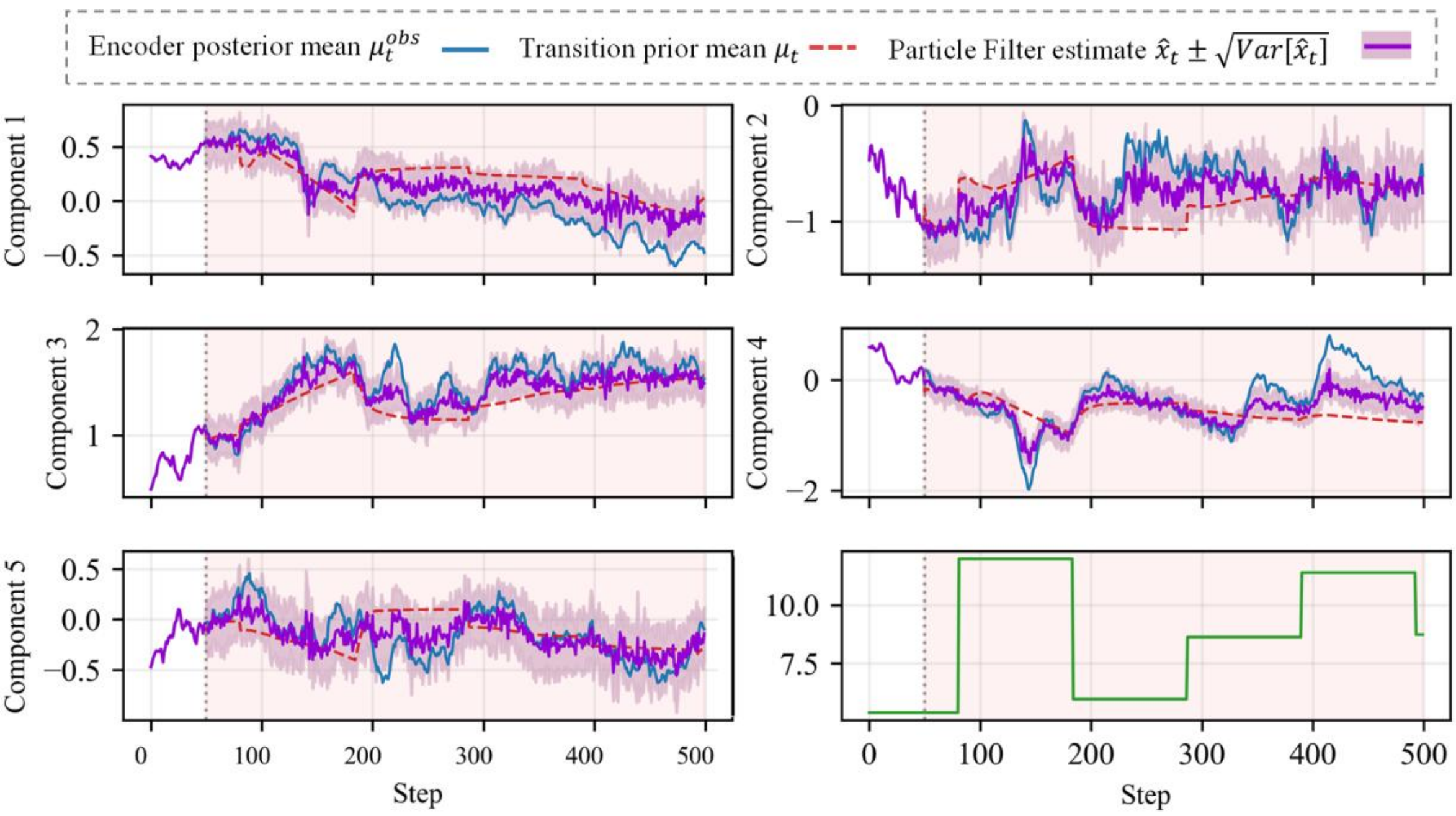

Fig. 14. Particle Filter State Estimation Results in GMAW.

## 6. Discussion and Future Work

The proposed DVAE–PF framework extracts compact, temporally coherent latent representations from high-dimensional weld-pool images and uses Particle Filtering (PF) to enhance robustness during deployment. Unlike conventional welding monitoring approaches relying on specialized architectures, handcrafted features, and large labeled datasets, our method offers two key advantages:

(1) Generalizability. Although demonstrated on arc-welding images, the framework is modality-agnostic. The DVAE's encoder–decoder can also be adapted to electrical signals [51], acoustic emissions [52], or multi-modal fusion [53], with an information-bottleneck design and divergence-based training ensuring meaningful low-dimensional latent states across heterogeneous data.

(2) Robustness. By modeling and enforcing causal relationships between process parameters and latent states, the framework learns temporally consistent latent dynamics that are resistant to noise and disturbances. The PF further enhances estimates by fusing latent dynamics with real-time observations, which is critical for industrial deployment under variable conditions.

We also envision the framework to be constructively beneficial in two aspects:

(1) Process monitoring. The PF-derived process state $\hat{x}_t$ can characterize the process in real time. The desired set of process states $x^\star$ can be obtained by encoding manually selected high-quality data frames. Thus, the process deviation can be quantified directly by the difference between $\hat{x}_t$ and $x^\star$.

(2) Model-based process optimization and control. Accurate process models, which describe the correlation between process parameters/control variables and process outputs, are essential for model-based design. The DVAE in our framework provides a data-enabled efficient approach to model process dynamics. The resultant model provides an effective surrogate of the underlying complex process dynamics to enable model-based control.

Overall, the DVAE–PF framework provides a unified and unsupervised approach to process monitoring and establishes a practical foundation for data-driven optimization and adaptive control in arc welding. Future work will expand the framework to heterogeneous sensing modalities, evaluate its performance in real-time monitoring, and develop closed-loop control strategies that exploit the learned latent dynamics.

## 7. Conclusions

This paper introduced an unsupervised probabilistic framework for robust state estimation in dynamic arc welding processes. By integrating deep latent representation learning with Bayesian filtering, the method addresses the critical challenge of inferring meaningful process states—such as weld penetration—from high-dimensional and noisy weld pool imagery in an unsupervised manner. The main contributions and findings are summarized as follows.

(1) A Dynamic Variational Autoencoder (DVAE) was developed to learn a compact, temporally coherent latent representation of weld pool dynamics. By unifying a CNN-based encoder–decoder with an LSTM-based transition model, the DVAE captures both the spatial semantics of individual weld pool images and their temporal evolution under process inputs, thereby overcoming the limitations of static representation learning.

(2) A specialized Particle Filter (PF) was designed to perform robust online state estimation in the DVAE latent space. Its hybrid stochastic–deterministic propagation scheme effectively fuses model-based predictions from the LSTM transition prior with observation-based encoder posteriors, producing smooth, accurate, and uncertainty-aware state estimates that are resilient to process noise and model drift.
(3) The framework demonstrated strong generalizability by being successfully applied to two widely used arc welding processes, GTAW and GMAW, without specific customization. This highlights the method's scalability and versatility for state inference across diverse welding applications.
In conclusion, the study establishes a unified paradigm for unsupervised, data-driven state estimation in manufacturing processes with complex dynamics. Future research will focus on integrating multi-sensor data, such as thermal and acoustic signals, to enrich latent state representations, and on leveraging the inferred states for real-time model-based feedback control, paving the way toward autonomous welding with guaranteed quality.

**Declaration:** This manuscript benefited from editorial refinement using ChatGPT (OpenAI), under the authors' direction and with full review and approval by the authors. The tool was used to improve clarity and readability; the authors take full responsibility for the content.

## Appendix A.

A central challenge in PF design is how to handles the LSTM's internal states $(h_{t-1}, c_{t-1})$, which encode the history of the welding process. While they are essential for a complete description of the dynamics, sampling them jointly with the latent state, as in a conventional particle filter, would not only impose heavy computational cost but also risk disrupting temporal consistency between the latent trajectory and the LSTM' internal states. To overcome this, we propose a hybrid stochastic–deterministic propagation scheme that strategically separates the LSTM's internal states from the latent state in particle propagation.

The PF algorithm, summarized in Table A.2, follows a recursive predict–update cycle, with additional resampling and estimation steps to ensure stability and interpretability:

(1) Prediction: Each particle is advanced using the learned LSTM dynamics. The internal states $(h_t^{(i)}, c_t^{(i)})$ are updated deterministically, while the transition prior outputs Gaussian parameters $(\mu_t^{(i)}, \Sigma_t^{(i)})$. A new latent state sample $x_t^{(i)} \sim \mathcal{N}(\mu_t^{(i)}, \Sigma_t^{(i)})$ is then drawn, introducing randomness to reflect process noise and unmodeled disturbances.

(2) Update: The latest observation $z_t$ is encoded by the DVAE into a posterior distribution $q_\phi(x_t|z_t) = \mathcal{N}(\mu_t^{\text{obs}}, \Sigma_t^{\text{obs}})$. Each particle's weight is updated according to how likely its predicted state is under this observation model. After normalization, the weights form a valid probability distribution, emphasizing particles consistent with the observation.

(3) Resampling: Over time, weights may collapse so that only a few particles dominate, a phenomenon known as degeneracy [33]. The Effective Sample Size (ESS) quantifies this effect. When $\text{ESS}_t$ falls below a threshold $\tau N$, resampling is performed to discard low-weight particles and replicate high-weight ones. This maintains diversity while concentrating particles in high-probability regions.

(4) Estimation: The fused state estimate $\hat{x}_t$ is obtained as the weighted mean of all particles, providing the best single estimate of the latent state. The associated variance $\text{Var}[\hat{x}_t]$ quantifies the confidence of this estimate, offering a principled uncertainty measure.

Table A.1. Notation for the Particle Filter.

| Symbol | Meaning |
|---|---|
| *States and Observations* | |
| $x_t$ | Latent state vector |
| $z_t$ | Observation (weld pool image) |
| $u_t$ | Control input (e.g., welding speed) |
| $h_t, c_t$ | LSTM hidden and cell states |
| $\mathcal{P}_t = \{(x_t^{(i)}, w_t^{(i)})\}_{i=1}^N$ | Particle set (state and weight) |
| $\hat{x}_t,\ \mathrm{Var}[\hat{x}_t]$ | PF state estimate and variance |
| *DVAE Distributions* | |
| $q_\phi(x_t\|z_t) = \mathcal{N}(\mu_t^{\mathrm{obs}}, \Sigma_t^{\mathrm{obs}})$ | Encoder posterior (Gaussian) |
| $p_{\theta_{\mathrm{trans}}}(x_t\|u_{t-1}, x_{t-1}, h_{t-1}, c_{t-1}) = \mathcal{N}(\mu_t, \Sigma_t)$ | LSTM transition prior (Gaussian) |
| $p_{\theta_{\mathrm{dec}}}(z_t\|x_t)$ | Decoder likelihood |
| *Filter Parameters* | |
| $\mathrm{ESS}_t$ | Effective Sample Size |
| $\tau$ | Resampling threshold (e.g., 0.5) |

Table A.2. Particle filter recursion algorithm.

| Symbol | Meaning |
|---|---|
| 1. Prediction | For each particle $i$:<br>- Update LSTM state: $h_t^{(i)}, c_t^{(i)} = \mathrm{LSTM}(u_{t-1}, x_{t-1}^{(i)}, h_{t-1}^{(i)}, c_{t-1}^{(i)})$<br>- Compute transition prior parameters: $\mu_t^{(i)}, \Sigma_t^{(i)} = p_{\theta_{\mathrm{trans}}}(u_{t-1}, x_{t-1}^{(i)}, h_t^{(i)}, c_t^{(i)})$<br>- Sample new state: $x_t^{(i)} \sim \mathcal{N}(\mu_t^{(i)}, \Sigma_t^{(i)})$ |
| 2. Update | - Encode observation: $(\mu_t^{\mathrm{obs}}, \Sigma_t^{\mathrm{obs}}) \leftarrow q_\phi(x_t\|z_t)$<br>- Update weights: $\tilde{w}_t^{(i)} = w_{t-1}^{(i)} \cdot \exp\left(-\frac{1}{2}(x_t^{(i)} - \mu_t^{\mathrm{obs}})^\top (\Sigma_t^{\mathrm{obs}})^{-1} (x_t^{(i)} - \mu_t^{\mathrm{obs}}) - \frac{1}{2}\log\|2\pi\Sigma_t^{\mathrm{obs}}\|\right)$<br>- Normalize: $w_t^{(i)} = \tilde{w}_t^{(i)} / \sum_j \tilde{w}_t^{(j)}$ |
| 3. Resample | Compute: $\mathrm{ESS}_t = 1/\sum (w_t^{(i)})^2$<br>If $\mathrm{ESS}_t < \tau N$:<br>- Resample particles $\{x_t^{(i)}\}$ with replacement proportional to $w_t^{(i)}$<br>- Reset weights: $w_t^{(i)} = 1/N$ |
| 4. Estimation | - $\hat{x}_t = \sum_{i=1}^N w_t^{(i)} x_t^{(i)}$<br>- $\mathrm{Var}[\hat{x}_t] = \sum_{i=1}^N w_t^{(i)} (x_t^{(i)} - \hat{x}_t)(x_t^{(i)} - \hat{x}_t)^\top$ |

.